\documentstyle[12pt,aaspp4,psfig]{article}
\eqsecnum
\received{}
\accepted{}
\journalid{}{}
\articleid{}{}
\slugcomment{Accepted for publication in {\it The Astronomical Journal}}

\begin{document}

\title{Observations of the core of the Pleiades with the Chandra X-ray
Observatory}

\author{Anita Krishnamurthi, Christopher S. Reynolds,\altaffilmark{1}
Jeffrey L. Linsky} 
\affil{JILA, University of Colorado and National Institute of Standards
and Technology, Boulder, CO 80309-0440; anitak@casa.colorado.edu,
chris@rocinante.colorado.edu, jlinsky@jila.colorado.edu}
\author{Eduardo Mart\'\i n}
\affil{Division of Geological and Planetary Sciences, California
Institute of Technology, MS 150-21, Pasadena, CA 91125;
ege@gps.caltech.edu}  
\author{Marc Gagn\'{e}}
\affil{Geology/Astronomy, 232 Boucher Hall, West Chester
University, West Chester, PA 19383; mgagne@wcupa.edu }

\altaffiltext{1}{Hubble Fellow}

\begin{abstract}

We present results from a 36-ksec observation of the core of the
Pleiades open cluster using ACIS-I on the Chandra X-ray Observatory.  We
have detected 57 sources, most of which do not have previously known
optical counterparts.  Follow-up photometry indicates that many of the
detections are likely to be AGNs, in accordance with extragalactic
source counts, but some of the sources may be previously undiscovered
low-mass members of the Pleiades.  We discuss our dataset and our
findings about X-ray emission from early-type stars as well as very late
type stars.  In particular, the large X-ray fluxes, lack of variability,
and hardness ratios of the four Pleiades B6 IV -- F4 V stars suggest a
tentative conclusion that Pleiades stars in this spectral type range are
intrinsic X-ray sources rather than previously unknown binaries in which
the X-ray emission is from a late-type companion. Also the sensitivity
of Chandra allowed us to detect nonflare X-ray emission from late-M
stars.


\end{abstract}

\section{Introduction}

Open clusters are an excellent laboratory for studying the evolution of
stellar properties as each cluster contains a sample of stars of the
same age and similar chemical composition formed under (presumably)
similar initial conditions.  Thus a comparison of clusters of different
ages allows us to study the evolution of angular momentum as well as
coronal and chromospheric activity as a function of age and mass.
Previous X-ray surveys, especially ROSAT, have shed light on coronal
properties in a large number of open clusters spanning a wide range of
ages (see Micela et al. (1999) for a comprehensive list).  Several of
these clusters also have measurements of rotation (in the form of
rotational velocities, $v\sin i$, or rotational periods, $P_{rot}$) for
a large sample of stars.

One important result of the ROSAT observations of these clusters is that
once the pre--main-sequence proto-planetary disk has been shed, F, G,
and K-type coronal activity, as evidenced by X-ray surface flux or
$L_{\rm X}/L_{\rm bol}$, is directly related to rotation rate; more
specifically it is related to the Rossby number ($N_{R}$, the ratio of
the rotation period to the convective overturn time; see review by
Jeffries (1999)).  Thus the rotation-activity connection has been
intensively studied this past decade across a range of masses and ages
in several open clusters, for e.g., IC2391/2602 (30 Myr;
\cite{patten96}, \cite{randich95}), $\alpha$ Per (50 Myr;
\cite{prosser96}), Pleiades (100 Myr; \cite{stauffer94};
\cite{micela99}), Hyades (670 Myr; \cite{stern95}).  We note that the
ages for these open clusters are not certain as other indicators (such
as the lithium depletion boundary method, e.g., \cite{stauffer00})
suggest that the age scale may need to be shifted to be systematically
older.  It is not yet clear whether the same
age-rotation-activity-paradigm extends to the fully convective M dwarfs
and the very young T Tauri stars, and it is completely unknown whether
it extends to the substellar brown dwarfs (BDs).

It was long thought that there might be a break in coronal properties at
the fully convective boundary, as one progresses from stars with
convective cores and radiative envelopes to fully convective stars.
Such an abrupt transition in chromospheric, transition-region, and
coronal activity has not been observed (e.g., \cite{fleming93}).  Hence,
the question arises whether the break is at the hydrogen burning mass
limit instead, i.e., BDs with $M<0.075 M_{\odot}$.

Several deep optical and near-infrared surveys have discovered very-low
mass stars and BDs in many star forming regions and young open clusters
(e.g., \cite{zap97}; \cite{luhman99}; \cite{stauffer99};
\cite{lucas00}).  X-ray emission has also been discovered from some of
these very low mass objects, including some BDs and from a larger sample
of brown dwarf candidates (BDCs; \cite{neuhauser99}).  We note that
persistent X-ray emission from BDs has been observed only in very young
objects (age $\sim$ 1 Myr).  Some X-ray flares have been seen from
late-type older objects such as VB 8 (spectral type M7;
\cite{giampapa96}) and VB 10 (spectral type M8; \cite{fleming00}).  Both
these sources have ages of the order of several times $10^{8}$ years
(\cite{reid94}).  Most recently, an X-ray flare was detected using the
Chandra X-ray observatory from the nearby brown dwarf LP 944-20 (age 500
Myr; \cite{rutledge00}).  None of these older sources have been detected
in X-rays outside of the flares.

Thus, there are several interesting questions to be answered with new
X-ray observations.  The Pleiades open cluster is ideal for studying
X-ray activity as a function of rotation and mass because (i) it is a
young (100 Myr), rich open cluster at a distance of only 125 pc, (ii) it
is one of the best studied open clusters, (iii) rotational
velocities or rotation periods have been measured for a large number of
stars spanning a wide range of mass and rotation rate (e.g.,
\cite{soder93}; \cite{anita98}; \cite{queloz98}), and (iv) the Pleiades
has long been a hunting ground for BDs at optical and infra-red
wavelengths (e.g., \cite{hambly93}; \cite{zap97}; \cite{festin97};
\cite{bouvier98}; \cite{zap99}; \cite{pinfield00}).
 
There have been several X-ray studies of the Pleiades conducted using
ROSAT (for e.g., \cite{stauffer94}, \cite{gagne95}, \cite{micela96},
\cite{micela99}).  In addition to providing us with a valuable X-ray
database for the Pleiades, these studies have shed light on several
important questions such as the rotation-activity paradigm for F, G and
K dwarfs, and luminosity functions for the earlier-type stars as well as
short-term and long-term variability.  However, while these surveys did
detect some early M stars, they did not detect many late-M stars due to
sensitivity limits.  Additionally, many BDs in the Pleiades have only
been discovered and cataloged recently.

In order to determine whether X-ray emission can be seen at the very
bottom of the main sequence and beyond, we observed the core of the
Pleiades cluster with the ACIS-I instrument aboard the Chandra X-ray
Observatory.  The much improved spatial resolution and higher effective
area of ACIS-I over the ROSAT PSPC provided an opportunity to detect the
small number of F, G, K, and early-M cluster members not detected by
ROSAT.  The ACIS-I images will also enable us to determine whether
persistent X-ray emission can be seen from late-M dwarfs and BDs older
than a few million years, and perhaps let us study the rotation-activity
relation across the H-burning boundary.

Section 2 discusses our observations and data reduction methods, while
section 3 discusses the main results.  Section 4 presents the
conclusions and future work.

\section {Data Acquisition and Reduction}

The observations presented in this paper were obtained as part of the
Guaranteed Time Observation (GTO) program for the Chandra X-ray
Observatory.  We were allocated 60 ksecs of observing time, which was
split into two parts.  This paper presents the first dataset comprising
36 ksec of data and the remaining data will be presented in paper 2.

The observation was centered on the core of the Pleiades cluster
(RA(J2000): 03h 46m 46s, Dec.(J2000): +24d 04m 06s) and was obtained
with the ACIS-I imaging array on September 18, 1999.  The imaging array
is comprised of four charge-coupled devices (CCDs) that are front-side
illuminated (I0, I1, I2, and I3).  All four chips of the imaging array
were active during this observation, giving us a square field
16$\arcmin$ by 16$\arcmin$ in size.  At the distance of the Pleiades,
this corresponds to a field of view 0.60\,pc $\times$ 0.60\,pc.  Two of
the chips from the ACIS-S array (S3 and S4) were also active during this
observation, but they were significantly out of the focal plane and
hence data from those chips were not analyzed for this paper.  The ACIS
detector was operated in graded mode throughout this observation, and
the standard 3.3\,s readout mode was employed.

The data reduction and analysis was done using the Chandra Interactive
Analysis of Observations (CIAO) software, developed by the Chandra
Science Center.  We started the analysis from the Level 2 events list
provided by the Chandra Science Center and retained only ASCA grades 0,
2, 3, 4, and 6.  Inspection of the resulting image shows bright stripes
on all front illuminated chips (including S4).  Formation of an image in
chip coordinates (rather than sky coordinates in which the dithering of
the space-craft has been accounted for) reveals that these stripes are
due to isolated `bad' columns on the chip.  Ignoring these bad columns
successfully eliminates the stripes with negligible loss of data.  We
also examined the time dependence of the background by making an `image'
in the energy-time plane.  Periods of high background, corresponding to
stripes in this energy-time image, were flagged and excluded from our
final clean image.

Using these methods, we produced a final clean image with a low
background level and free of defects.  Figure 1 shows a smoothed ACIS-I
image of the Pleiades obtained in 36 ksecs.  The image was generated
using the CSMOOTH algorithm in CIAO.  This algorithm uses Fourier
techniques to perform adaptive smoothing of the image such that any
structure is smoothed out unless it is present above some predetermined
level of significance.  We chose to smooth to the $4\sigma$ level.
Numerous point sources can be seen in this image.  The diffuse structure
seen in the background is likely due to the smoothing of the
instrumental effects and diffuse cosmic background.

Data from the front-side illuminated (ACIS-I) CCDs in our dataset are
affected by the degradation of charge transfer efficiency in the
low-energy range 0.2-0.5 keV, caused by the high dosage of particle
irradiation early in the Chandra mission (\cite{prigozhin00}).  A charge
transfer inefficiency (CTI) corrector has been developed to correct this
problem (\cite{town00}).  However, as our data were obtained in
3$\times$3 ``graded mode'' rather than in 3$\times$3 ``faint mode'', the
3$\times$3 event neighborhood was not telemetered to the ground.  As
these neighborhoods are required by the corrector to reconstruct the
CTI-free event, we are not able to correct our dataset for this problem.

The primary goal of this work is to identify the point sources in this
field.  Initially, we attempted to use the CIAO routine ``celldetect'',
which is a simple detection algorithm based on the Poisson statistics
within a sliding box.  We found that, near the center of the field, this
routine missed sources which were clearly present upon a visual
inspection.  Instead, source detections were obtained using the
wavelet-based algorithm ``wavdetect'' in CIAO.  We set the false alarm
probability set to 10$^{-6}$ per pixel.  This implies that there would
be 1 spurious source detected per 1000$\times$1000 pixel image.  Our
binned image was 512$\times$512 pixels, so we would expect 0.25 spurious
sources in our field.  We have detected 57 sources using this method.
Careful visual inspection of the field gives us confidence that
``wavdetect'' has performed its task successfully.

We also ran the ``wavdetect'' algorithm with the false alarm probability
set to 10$^{-5}$ per pixel.  This lower threshold picked up 15
additional sources, listed in Table 3.  From an inspection of the X-ray
image, we believe it is likely that most of these sources are not real.

\section{Results and discussion}

{\tiny
\begin{deluxetable}{lllllcc}
\tablenum{1}
\tablewidth{0pt}
\pagestyle{empty}
\tablecaption{Sources detected by Chandra.}
\tablehead{
\colhead{Number} & 
\colhead{RA (2000)}& 
\colhead{DEC (2000)} &
\colhead{Counts} & 
\colhead{Error} &
\colhead{Hardness}&
\colhead{Error in} \\
\colhead{} &
\colhead{} &
\colhead{} &
\colhead{} &
\colhead{} &
\colhead{ratio (HR)} &
\colhead{HR}
}
\startdata
1  & 3 47 25.25  &  24  2 54.10   &   75.15   &   11.23  &  -0.39   &   0.14 \\
2  & 3 47 18.10  &  24  2 11.17   &  785.24   &   29.21  &  -0.11   &   0.04 \\
3  & 3 47 16.98  &  24 12 33.70   &   50.67   &    9.00  &   0.59   &   0.09 \\
4  & 3 47 16.48  &  24  7 41.84   &  184.59   &   15.14  &  -0.56   &   0.08 \\
5  & 3 47 14.10  &  24  3 18.70   &   12.94   &    5.10  &  -0.11   &   0.37  \\
6  & 3 47  9.11  &  24  3  7.56   &  189.64   &   14.70  &  -0.20   &   0.08  \\
7  & 3 47  7.59  &  24 10 16.19   &  111.14   &   12.53  &   0.80   &   0.02 \\
8  & 3 47  5.71  &  23 59 42.30   &   40.85   &    9.06  &   0.63   &   0.08  \\
9  & 3 47  3.57  &  24  9 34.86   &  292.16   &   18.11  &  -0.33   &   0.06  \\
10 & 3 47  1.66  &  24 10 47.53   &   28.95   &    7.28  &   0.91   &   0.02  \\
11 & 3 47  1.29  &  24  2  6.07   &   29.11   &    7.07  &   0.45   &   0.17  \\
12 & 3 47  0.80  &  24 13 24.56   &   20.22   &    6.56  &   0.46   &   0.18  \\
13 & 3 46 59.25  &  24  1 42.54   &   85.48   &   10.15  &   0.05   &   0.11  \\
14 & 3 46 58.15  &  24  1 40.39   &   41.70   &    7.68  &  -0.43   &   0.19  \\
15 & 3 46 55.74  &  24  1 52.67   &   27.99   &    6.56  &   0.52   &   0.09  \\
16 & 3 46 55.63  &  24  2 56.88   &   20.26   &    5.75  &   0.25   &   0.22  \\
17 & 3 46 55.17  &  24 11 14.27   &   27.56   &    6.86  &  -0.30   &   0.38  \\
18 & 3 46 54.06  &  24  7 56.49   &   12.74   &    5.10  &  -0.28   &   0.31  \\
19 & 3 46 53.05  &  23 56 55.20   &   63.53   &    9.44  &   0.74   & 0.04  \\
20 & 3 46 51.97  &  24  0  4.61   &   29.09   &    6.78  &   0.33   &   0.13  \\
21 & 3 46 51.31  &  24  6 16.17   &   24.04   &    6.41  &  -0.17   &   0.23  \\
22 & 3 46 49.13  &  24  8 33.45   &   40.33   &    7.62  &   0.30   &   0.12  \\
23 & 3 46 48.54  &  24  0 40.57   &   21.44   &    6.09  &   0.52   &   0.10  \\
24 & 3 46 47.23  &  24  1 28.87   &   81.81   &   10.25  &   0.40   &   0.08  \\
25 & 3 46 46.71  &  24  7 57.58   &   42.13   &    7.55  &   0.45   &   0.11  \\
26 & 3 46 44.29  &  23 59 11.53   &   18.54   &    6.09  &   0.38   &   0.16  \\
27 & 3 46 43.52  &  23 59 41.51   &  109.11   &   11.23  &  -0.21   &   0.11  \\
28 & 3 46 41.94  &  24  3 51.17   &   18.67   &    5.75  &   0.25   &   0.19  \\
29 & 3 46 41.29  &  24  4 26.01   &   20.28   &    5.39  &   0.35   &   0.18  \\
30 & 3 46 40.99  &  24  7 46.32   &   26.35   &    6.41  &   0.75   &   0.05  \\
31 & 3 46 39.34  &  24  1 46.22   &  425.09   &   21.19  &  -0.24   &   0.11  \\
32 & 3 46 39.30  &  24  6 11.17   &  745.54   &   28.04  &  -0.51   &   0.04  \\
33 & 3 46 38.80  &  23 58  5.12   &   64.03   &    9.49  &   0.31   &   0.12  \\
34 & 3 46 35.84  &  23 58  0.39   &  209.48   &   15.33  &  -0.36   &   0.08  \\
35 & 3 46 35.66  &  24  7 52.78   &   33.41   &    7.07  &   0.57   &   0.10  \\
36 & 3 46 35.45  &  24  1 35.44   &   36.65   &    7.21  &  -0.60   &   0.17  \\
37 & 3 46 32.39  &  24  5 47.81   &   69.26   &    9.33  &   0.45   &   0.09  \\
38 & 3 46 32.06  &  23 58 58.15   &  789.31   &   28.90  &  -0.14   &   0.04  \\
39 & 3 46 31.11  &  24  7  1.74   &  280.49   &   17.61  &  -0.27   &   0.07  \\
40 & 3 46 30.32  &  24  8 53.80   &   31.43   &    7.68  &   0.40   &   0.17  \\
41 & 3 46 30.41  &  24  5  5.63   &   20.37   &    6.48  &  -0.19   &   0.28  \\
42 & 3 46 29.47  &  24  0 41.88   &   31.14   &    7.07  &   0.38   &   0.17  \\
43 & 3 46 28.15  &  23 57 54.75   &   37.42   &    7.81  &   0.25   &   0.17  \\
44 & 3 46 28.01  &  23 55 32.46   &   37.26   &    7.62  &   0.35   &   0.16  \\
45 & 3 46 25.30  &  24  9 36.06   &  336.23   &   19.29  &   0.00   &   0.06  \\
46 & 3 46 23.70  &  23 55 38.87   &  293.24   &   18.95  &   0.58   &   0.04  \\
47 & 3 46 23.36  &  24  1 51.15   &   49.90   &    8.49  &  -0.23   &   0.17  \\
48 & 3 46 23.39  &  24  8 56.02   &   15.40   &    5.48  &   0.64   &   0.08  \\
49 & 3 46 21.56  &  24  7 17.02   &   48.32   &    8.66  &   0.71   &   0.07  \\
50 & 3 46 20.85  &  23 58 58.52   &  123.08   &   12.57  &  -0.01   &   0.10  \\
51 & 3 46 19.52  &  23 56 53.05   & 1146.65   &   35.60  &  -0.33   &   0.03  \\
52 & 3 46 19.42  &  23 56 29.18   &   26.00   &    7.07  &   0.47   &   0.09  \\
53 & 3 46 17.71  &  24  1 10.51   &   32.96   &    8.07  &   0.46   &   0.11  \\
54 & 3 46 15.91  &  24 11 22.26   &  988.97   &   33.63  &  -0.48   &   0.04  \\
55 & 3 46 12.77  &  24  3 14.73   &  303.41   &   18.47  &  -0.19   &   0.06  \\
56 & 3 46 24.91  &  24  1 29.30   &   25.17   &    7.42  &   0.48   &   0.17  \\
57 & 3 46 17.45  &  23 58 32.80   &   23.63   &    7.00  &   0.77   &   0.05  \\
\enddata
\end{deluxetable}
}

The 57 sources detected by Chandra in the Pleiades field are listed in
Table 1.  The first column refers to the source number, columns 2 and 3
specify the positions, columns 4 and 5 list the counts and the
statistical error on the counts, and columns 6 and 7 list the hardness
ratios (discussed further in Section 3.2).  Figure 2 shows the ACIS-I
detections (little black boxes) overplotted on an optical image of the
Pleiades field of view.  The optical image used is from the ``skyview''
database (operated by NASA/GSFC), which incorporates the ``Digitized Sky
Survey''.  Note that we do not see optical counterparts for several
X-ray detections in this image and conversely, we do not have X-ray
detections for several optically bright stars in the field.

{\small
\begin{deluxetable}{llllllll}
\tablenum{2}
\tablewidth{0pt}
\pagestyle{empty}
\tablecaption{Known optical counterparts to ACIS detections.}
\tablehead{
\colhead{Number} & 
\colhead{Optical} &
\colhead{Offset} &
\colhead{Corrected} &
\colhead{X-ray} & 
\colhead{V}&
\colhead{Spectral}&
\colhead{Member}\\
\colhead{} &
\colhead{counterpart} &
\colhead{($\arcsec$)} &
\colhead{offset($\arcsec$)} &
\colhead{counts} &
\colhead{} &
\colhead{type} &
\colhead{}
}
\startdata
51 & hii 980   & 3.3 & 1.28 & 1147 & 4.18  & B6IV & Y \\
54 & hii 956   & 3.0 & 1.10 & 989 & 7.96   & A7V  & Y \\
4  & hii 1338  & 2.0 & 0.15 & 185 & 8.69   & F3V  & Y \\
32 & hii 1122  & 2.7 & 1.00 & 745 & 9.29   & F4V  & Y \\
31 & hii 1124  & 1.9 & 0.33 & 425 & 12.32  & K1V  & Y \\
2  & hii 1355* & 1.8 & 0.51 & 785 & 14.02  & K6V  & Y \\
34 & hii 1094  & 2.1 & 0.38 & 209 & 14.02  & KV   & Y \\
55 & hii 930*  & 2.3 & 0.67 & 303 & 14.20  & KV   & Y \\
39 & hii 1061  & 2.1 & 0.16 & 280 & 14.21  & K5V  & Y \\
38 & hcg 225*  & 2.7 & 0.77 & 789 & 14.55  & --   & N \\
9  & hii 1280* & 1.4 & 1.11 & 292 & 14.57  & K7.5V& Y \\
47 & hcg 218   & 1.4 & 0.9  & 50  & 15.63  & --   & N \\
45 & hcg 219*  & 1.6 & 0.43 & 336 & 15.83  & KV   & Y \\
1  & hhj 427   & 3.4 & 1.42 & 75  & 16.10  & MV   & Y \\
13 & hhj 299*  & 1.1 & 1.11 & 85  & 17.60  & MV   & Y \\
18 & MHO-8     & 2.2 & 1.65 & 12  & 18.92  & MV   & ? \\
36 & hhj 140   & 0.9 & 2.1  & 37  & 19.00  & MV   & Y \\
17 & MHO-9     & 3.3 & 2.43 & 27  & 19.02  & MV   & ? \\
\tablenotetext{}{Hii refers to the second list in Hertzsprung
(1947), sometimes also referred to as Hz; HCG refers to the catalog of
Haro, Chavira \& Gonzalez (1982); HHJ is the Hambly, Hawkins and Jameson
catalog (1991); MHO objects are from the Mount Hopkins survey (Stauffer
et al. 1998).}
\tablenotetext{}{* indicates a star has flared during the observation.
Discussion in sections 3.4 and 3.5.} 
\tablenotetext{}{? indicates uncertain membership.}
\enddata
\end{deluxetable}
}

We checked for optical counterparts to our detections by comparing
against a compiled master list of sources in the Pleiades, including
recently conducted deep surveys to detect brown dwarfs.  The optical
counterparts which lie within 5$\arcsec$ of the X-ray detections are
presented in Table 2.  We also tested for counterparts within
10$\arcsec$ to see if we picked up any additional sources, but we did
not.  Figure 3 shows a plot of offsets between the Chandra detections
and the optical positions of the well-cataloged Pleiades members.  The
largest offset is 3.3$\arcsec$.  Since the position offsets are mostly
in the same direction, we believe that the pipeline-processed Chandra
aspect solution could be in error and therefore obtained a least squares
correction to the aspect solution.  The best fit is obtained if we shift
the X-ray positions by $\Delta$RA = --0.60$\arcsec$ and $\Delta$Dec =
+1.88$\arcsec$.  The corrected offsets are listed in column 4 of Table
2.  The correction reduces the mean offset from 2.18$\arcsec$ to
0.97$\arcsec$.

In some cases, the Chandra image resolved sources previously thought to
be single (from ROSAT data) into multiple sources.  This will be
discussed further in an upcoming paper.  Note that only 18 of the
sources detected by Chandra can be definitely correlated with optical
counterparts from our master list.  Although there have been several
deep surveys in the Pleiades (e.g., \cite{zap99}; \cite{bouvier98})
these typically have not included the core of the cluster where our
observations lie.

\subsection{Keck photometry}

In order to determine whether the X-ray sources without any optical
counterparts are very low mass members of the Pleiades, we obtained some
follow-up photometry of these fields.  We obtained CCD images at the
Keck II telescope using the Low Resolution Imaging Spectrograph (LRIS;
\cite{oke95}) on January 6, 2000.  Three fields were observed, centered
at 03:46:55, +24:07:00; 03:46:46, +23:59:00; and 03:46:24, +24:06:00
(J2000).  The field of view of LRIS is 6$\times$7.8$\arcmin$, with a
scale of 0.215 arcsec/pix.  Thus, we covered 140.4 arcmin$^{2}$, or
about 60\% of the area of our Chandra image.  For each field we obtained
two exposures with the standard broadband filter R, with exposure times
90 and 600 seconds; and two exposures in the I-band, with exposure times
of 60 and 300 seconds.  Stars brighter than R=16.9 and I=16.5 are
saturated in our short exposures.  Our detection limit for point sources
(5~$\sigma$) is R=24.4, I=23.5.

These observations were bracketed by images of the cool star CFHT-Pl-20,
for which R, I photometry is available in Bouvier et al. (1998).
Mart\'\i n et al. (2000) have shown that CFHT-Pl-20 is a low-mass field
star, and not a Pleiades member.  We used CFHT-Pl-20 as a standard to
calibrate our data. Thus, our R and I photometry is in the same system
as that of Bouvier et al. (1998).  We reduced the data using standard
IRAF/DIGIPHOT routines.  We performed point spread function (PSF)
photometry of all the sources detected with DAOFIND.  The FWHM of the
PSF ranged from 2.1 to 3.6 pixels (0.45$\arcsec$ to 0.77$\arcsec$),
depending on the position on the detector.  All the data was obtained at
airmass less than 1.15.  However, due to thin cirrus, our data are not
photometric.  We see scatter in the photometry of the stars of the
CFHT-Pl-20 field of up to 0.15 magnitudes in I-band and up to 0.07
magnitudes in R-band.  Adding the contribution of this scatter and the
photon-noise errors, the errors in our photometric measurements are
better than 0.2 mag., which is sufficient for our purposes of
identifying faint optical counterparts to the X-ray sources.

Figure 4 shows an I vs. R-I plot for the Keck sources that lie within
10$\arcsec$ of the Chandra detections.  From this figure, it is seen
that the Keck sources are too blue to be low mass Pleiades members.  We
were able to find optical counterparts for 21 of the 39 ``mystery
sources'', but the Keck photometry covers only about 60\% of the Chandra
field of view.  Some Chandra detections might indeed be low mass
Pleiades members as discussed in the next section.

\subsection{Hardness ratios}

Since the ACIS-I detectors have intrinsic energy resolution, we can
examine the spectral properties of the detected sources.  Since most of
the sources have insufficient counts to form a spectrum, we will use
hardness ratios.  As discussed below, this is an important constraint on
the nature of these sources.

Hardness ratios were computed using the following method.  Starting with
the level-2 events list, we formed a soft (0-1\,keV) and a hard
(1-3\,keV) image.  The critical values of the pulse height amplitude
(pha) corresponding to these energy bounds was determined using a
redistribution matrix file (rmf) derived from the appropriate FITS
embedded function (FEF) file.  Next, we used ``wavdetect'' to find all
of the sources in each image, using a very low false-alarm probability
(FAP) of 10$^{-4}$.  We used such a low FAP in order to detect as many
of the full-band sources as possible.  We then attempted to identified
our 57 cataloged sources in the soft and hard band images (with a
spatial tolerance of 3$\arcsec$), and measured their soft and hard band
count rates.  This procedure allowed the soft and hard band count rates
to be measured for all but 15 of the faintest sources.  For these 15
sources, we measured the soft and hard band count rates by manually
extracting and counting the photons from the events list using the
source position as determined from the full band image.  Note that the
numerous spurious `sources' found in the hard and soft band images due
to the low FAP of $10^{-4}$ are irrelevant since we base our
identifications on a much more rigorous search of the full band image.

Given the soft and hard band count rates, we define a hardness ratio for
each source as:
\begin{equation}
HR = \frac{hard - soft}{hard + soft}
\end{equation}
and compute the standard error in that ratio assuming Poisson statistics
in each band.  In order to use the hardness ratio to constrain the
nature of these sources, we have used XSPEC to compute HR for a few
canonical spectra.  We find that a standard AGN spectrum (with photon
index, $\Gamma=2$, and absorbed by the Galactic column density in the
direction of the Pleiades of N$_{H}$=1.1$\times$10$^{21}$ cm$^{-2}$,
\cite{dickey90}) would have HR$_{AGN}$ = +0.35.  In practice, luminous
AGN may be harder than this (i.e. $\Gamma<2$), and may have intrinsic
absorption.  Both of these effects will tend to make the source harder
(i.e. HR $>$ 0.35).  Ultrasoft AGN (such as narrow line Seyfert 1
galaxies; \cite{boller96}) can have photon indices as soft as
$\Gamma=4$, giving HR$_{SAGN}=-0.32$.  Hence, any source with HR$<-0.32$
is unlikely to be an AGN.  Indeed, no known type of extragalactic X-ray
source will be softer than this once the effects of the Galactic
absorption have been taken into account.  It is important to note that
the source counts of Mushotzky et al. (2000), corrected for the
relatively high column density along the Pleiades line of sight, imply
$\sim 35$ extragalactic sources in our field.

Given the strong soft X-ray absorption when looking out of our Galaxy,
any source softer than HR$<-0.32$ is likely to be a stellar source in
our Galaxy.  However, the converse is not true -- stellar sources may
well be harder than this critical threshold.  To be more quantitative, a
star with an X-ray spectrum characterized as a thermal plasma spectrum
(kT=0.5keV) absorbed by the column density to the Pleiades will have
HR$_{0.5keV}=-0.5$.  On the other hand, choosing $kT=0.2$\,keV or
$kT=2$\,keV gives HR$_{0.2keV}=-0.85$ and HR$_{2keV}=+0.5$ respectively,
thereby demonstrating that plausible stellar X-ray spectra can give a
wide variety of hardnesses ranging from very-soft to moderately hard.

Examining the hardness ratios for 16 Pleiades members, we see
that they are all rather soft, spanning the range from HR$=-0.60\pm
0.17$ to HR$=0.05\pm 0.11$.  There is a slight tendency for the bright
members with V$<$10 to be soft X-ray sources (Fig.~5a).  A histogram of
the hardness ratio for the Pleiades members and the other X-ray sources
shows a very clear demarcation --- the vast majority of the 41 X-ray
sources that are not identified with Pleiades members are hard.  In
addition, the Keck photometry available for some of these other sources
show that they are too blue to be Pleiades low mass members.  We suggest
that the ten sources with HR$>$0.5 are almost certainly background AGN,
probably with a large amount of intrinsic absorption.  These are exactly
the type of objects believed to make up the hard X-ray background
(\cite{comastri95}).  An additional 22 non-Pleiades sources have HR
consistent with the canonical AGN value HR$=0.32$ and are also likely to
be background AGN.  The remaining 9 non-Pleiades sources could be either
ultra-soft AGN or previously unknown low mass stars either in the
Pleiades or along the Pleiades line of sight.  In particular, we note
that the softest 4 of these sources have hardness ratios consistent with
HR$<-0.32$ and hence might be too soft to be extragalactic sources.
Unfortunately, more data is required in order to reduce the error bars
on HR and make this conclusion a firm one.

\subsection{Flux determination}

To determine fluxes for the Chandra detections, we started by extracting
spectra for the X-ray bright sources that have optical counterparts.
We then utilized the package ``XSPEC'' to fit a two-temperature Mekal
model assuming a hydrogen column density to the Pleiades of N$_{H}$ =
10$^{20.5}$ cm$^{-2}$ (\cite{cai85}), kT$_{1}$ was fixed to 0.544 keV
(\cite{gagne99}) and kT$_{2}$ was allowed to vary.  We determined the
corrected fluxes in the 0.5-2.0 keV range and found that the mean
counts-to-flux conversion factor was 1.55$\times$10$^{-16}$
erg~~s$^{-1}$~cm$^{-2}$.  Assuming a constant counts-to-flux conversion
factor, we derived fluxes for all our detections.  Figure 6 shows a
histogram of fluxes for the Chandra detections, the known Pleiades
members are shown as dashed lines.

Assuming a distance of 127 pc to the Pleiades to be consistent with
previous ROSAT measurements (\cite{stauffer94}), we determined the
luminosities for the known Pleiades members.  Although there are only a
few known Pleiades members in our sample, the lowest flux measured
corresponds to log(L$_{x}$)=27.5.  This is an order of magnitude fainter
than the comprehensive ROSAT survey carried out by Stauffer et
al. (1994).  We found that the difference between the ROSAT and Chandra
luminosities for the sources in common between our dataset and the ROSAT
survey varies from 10\% to a factor of 3.  We suspect that source
variability and the CTI effects may account for the difference.  This
issue will be explored further in paper 2.

\subsection{Coronal variability and the early-type stars}

Light curves were extracted for all 57 detected sources; we used the
CIAO tool ``dmextract'' to produce light curves binned in 2000 second
intervals for the sources with more than 100 counts, and binned the
sources with less than 100 counts more coarsely (8000 second bins).  We
assessed whether the stars had flared using two techniques: (i) a simple
visual examination of the light curves, and (ii) determining the
$\chi^{2}$ values resulting from fitting a constant model to the light
curve and finding those cases where the constant model is rejected at
the 99\% level (\cite{bev92}).  

Figure 7 shows a panel of representative light curves for some of the
stars in the sample: one B sub-giant, one A-star, one F-dwarf (none of
which are variable), a K dwarf that has flared, an M dwarf that has
flared and an M dwarf that is not variable.

A long standing question is whether A stars have coronae or whether the
X-ray emission sometimes seen from early-type stars is due to an unseen
later-type companion (\cite{grillo92}; \cite{schmitt93};
\cite{micela96}).  Most X-ray surveys of clusters with early-type
members find that only a small fraction of the sample emit X-rays.  This
has been interpreted as being consistent with the scenario of a low-mass
companion emitting X-rays.  While our limited data set does not allow us
to make a definitive statement concerning whether Pleiades age A-type
stars are intrinsic X-ray sources or the emission is from late-type
companion stars, the increased sensitivity of Chandra compared to ROSAT
does allow us to identify an interesting trend that should be pursued in
subsequent observations. With the more sensitive Chandra data, we can
bring source variability and X-ray hardness ratio data into the
analysis. The four ``early-type'' stars in the sample (spectral types
B6IV, A7V, F3V, and F4V) are not variable, have soft hardness ratios (HR
between --0.33 and --0.56), and have large X-ray fluxes (an average of
767 X-ray counts). By comparison, five of the eight K stars in our
sample are variable in X-rays. There is a large difference between the
five K stars that are variable (an average of 501 counts and HR between
--0.33 and +0.0) and the three K stars that are not variable (an average
of 180 counts and HR between --0.36 and --0.23).

With these data we now consider the hypothesis that the emission from
the sample of four ``early-type'' stars is due to putative K stars
companions.  If we assume that the X-ray emission from the four
``early-type'' stars is produced by K dwarf companions like the five
variable stars, then the mean X-ray count rates could be explained (767
counts for the ``early-type'' stars vs 501 for the variable K stars),
but the absence of X-ray variability and soft HRs of the ``early-type''
stars cannot be explained.  If, on the other hand, we assume that the
companion stars are like the three nonvariable K stars, then the average
X-ray flux of the early-type stars is 4.3 times too large (767 vs 180
counts). The only alternative left is that the B6IV--F4V stars are
intrinsic X-ray sources.  This admittedly unorthodox conclusion must be
considered tentative at this point as additional long Chandra
observations of many A-type stars are needed to confirm or refute the
hypothesis. Such observations are now underway.

\subsection{X-ray emission from very low mass members}

Convective, rapidly rotating stars have dynamo-generated magnetic fields
and hence coronal heating.  Stars later than a spectral type of M5 are
fully convective and thus cannot have the standard solar-like
($\alpha\Omega$) dynamos.  Instead, strong chaotic magnetic fields can
be generated by thermally driven turbulent convection without the need
for helicity or a lower boundary with a radiative core as likely occurs
just below the solar photosphere (\cite{cattaneo99}).  A survey of
late-M dwarfs conducted using ROSAT (\cite{fleming93}) showed that
late-M dwarfs also emit X-rays and thus must have hot coronae.
Therefore, it is fully expected that fully convective BDs, with or
without rapid rotation, should also be X-ray sources.  Very late
M-dwarfs such as VB 8 (spectral type M7; \cite{giampapa96}) and VB 10
(spectral type M8; \cite{fleming00}) have been seen only during flares
and only upper limits exist for their quiescent flux.  Also, X-ray
emission from brown dwarfs had been seen only in very young objects (age
$\sim$1 Myr, \cite{neuhauser99}) until recently.  Chandra has now
detected a flare on the relatively old (500 Myr) nearby field brown
dwarf LP 944-20 (\cite{rutledge00}).  We have detected X-ray emission
from M dwarfs (including some late-M stars) in the Pleiades.  It is
worth noting that of five possible M dwarf Pleiades members, (hhj 427,
hhj 299, hhj 140, MHO-8 and MHO-9) only one star has flared (hhj 299);
we have detected quiescent fluxes for the remaining objects.

There are also several potential BDCs seen in the Keck images obtained
for follow-up photometry of the Chandra detections with no optical
counterparts.  While the substellar nature of these objects needs to be
confirmed with follow-up spectroscopy, none of these BDCs have been
detected in our Chandra dataset.  We can put an upper limit of log
L$_{x} \sim$27.5 on X-ray emission from these objects if they are
confirmed to be Pleiades brown dwarfs.  We will be obtaining follow-up
spectroscopy of these potential brown dwarfs.  Results will be presented
and discussed in the succeeding paper.

\section{Conclusions}

We have presented results of a 36 ksec observation using ACIS-I on
Chandra of the core of the Pleiades cluster.  We found that most of our
detections did not have previously known optical counterparts and we
obtained follow-up photometry utilizing the Keck LRIS.  Several sources
seen in the Chandra dataset but with no previous IDs have been detected
in the Keck dataset.  Hardness ratio calculations indicate that many of
these sources are likely extra-galactic, but a small fraction are
soft and may be good candidates for Pleiades very low mass stars or
substellar objects.  We have also detected quiescent X-ray emission from
some late-M dwarfs in the Pleiades not previously seen in X-rays.

Some brown dwarf candidates have been identified in the Keck dataset.
Despite ACIS-I's great sensitivity, these BDCs have not been detected in
X-rays.  If planned follow-up spectroscopy of these objects confirms
them to indeed be brown dwarfs, this lack of X-ray emission from brown
dwarfs in the Pleiades will shed light on the role of coronal heating
and magnetic fields in older brown dwarfs.

A subsequent paper will present the results of the remaining 24 ksecs of
data and present the Keck photometry with possible IDs for the sources,
as well as the results of any follow-up spectroscopy.  Additionally, as
these two datasets were taken 6 months apart, we will be able to study
the X-ray variability of the detected objects.

\acknowledgements

We gratefully acknowledge the help provided to us with compiling the
Pleiades database by John Stauffer, G. Micela, and J. Adams.  We are also
very grateful to Leisa Townsley for helpful discussions regarding the
CTI effects in our dataset.  AK also thanks the Chandra Science Center
and the ``helpdesk'' for the help received during reduction of this
dataset.  AK and JLL acknowledge support from grant number H-04630D to
the University of Colorado and NIST.

CSR appreciates support from Hubble Fellowship grant HF-01113.01-98A.
This grant was awarded by the Space Telescope Institute, which is
operated by the Association of Universities for Research in Astronomy,
Inc., for NASA under contract NAS 5-26555.  CSR also appreciates support
from the National Science Foundation under grants AST-9529170 and
AST-9876887.

\section*{Appendix}

We present a table of the 15 additional sources picked up by
``wavdetect'' if FAP is set to 10$^{-5}$ rather than 10$^{-6}$.

\begin{deluxetable}{lccc}
\tablenum{3}
\tablewidth{0pt}
\tablecaption{Additional sources detected with FAP=10$^{-5}$}
\tablehead{
\colhead{RA(J2000)} &
\colhead{Dec(J2000)} &
\colhead{X-ray counts} & 
\colhead{Error}
}
\startdata
3 47 11.70  & 24  6 53.58  &  12.89  & 4.69 \\
3 47  4.02  & 23 59 42.68  &   8.41  & 4.13 \\
3 47  3.30  & 24 12 14.07  &   8.79  & 4.47 \\
3 46 59.54  & 24  4 21.01  &  15.93  & 5.83 \\
3 46 49.03  & 24  0  3.47  &  12.24  & 5.29 \\
3 46 48.08  & 24  3 34.63  &   8.22  & 4.13 \\
3 46 43.64  & 24  1 20.57  &   7.70  & 4.13 \\
3 46 34.10  & 24  1 49.53  &  20.95  & 7.00 \\
3 46 12.70  & 24  3 45.38  &  20.67  & 6.48 \\
3 46  9.84  & 24  9 54.32  &  23.45  & 6.41 \\
3 47 22.49  & 24  0  9.89  &  18.50  & 6.17 \\
3 46 57.40  & 24 13  7.10  &  13.09  & 5.39 \\
3 46 46.01  & 24  5  3.26  &  21.74  & 6.86 \\
3 46 15.95  & 24  8 26.20  &   9.51  & 4.24 \\
3 46  7.15  & 24 11  5.91  &   9.51  & 3.87 \\
\enddata
\end{deluxetable}

\clearpage

\begin{figure}
\centerline{
\psfig{figure=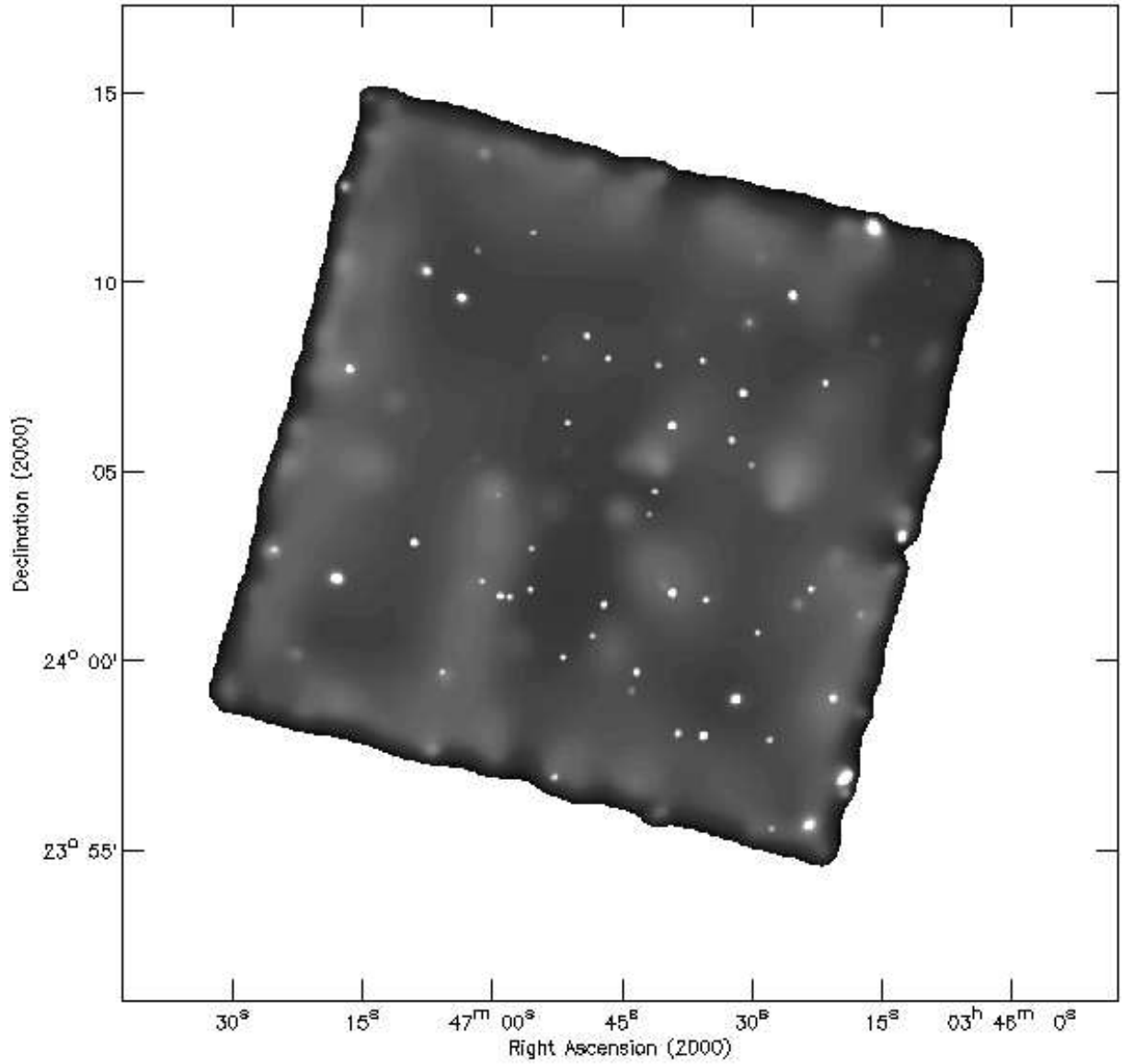,width=1.0\textwidth}
}  
\caption{Smoothed ACIS-I image of the Pleiades obtained with Chandra in
36 ksecs.  This image was generated using the CSMOOTH algorithm in CIAO
1.0.  This routine uses a variable smoothing length, such that any
structure is smoothed out unless it is significant at the 4$\sigma$
level.}
\end{figure}

\begin{figure}  
\centerline{
\psfig{figure=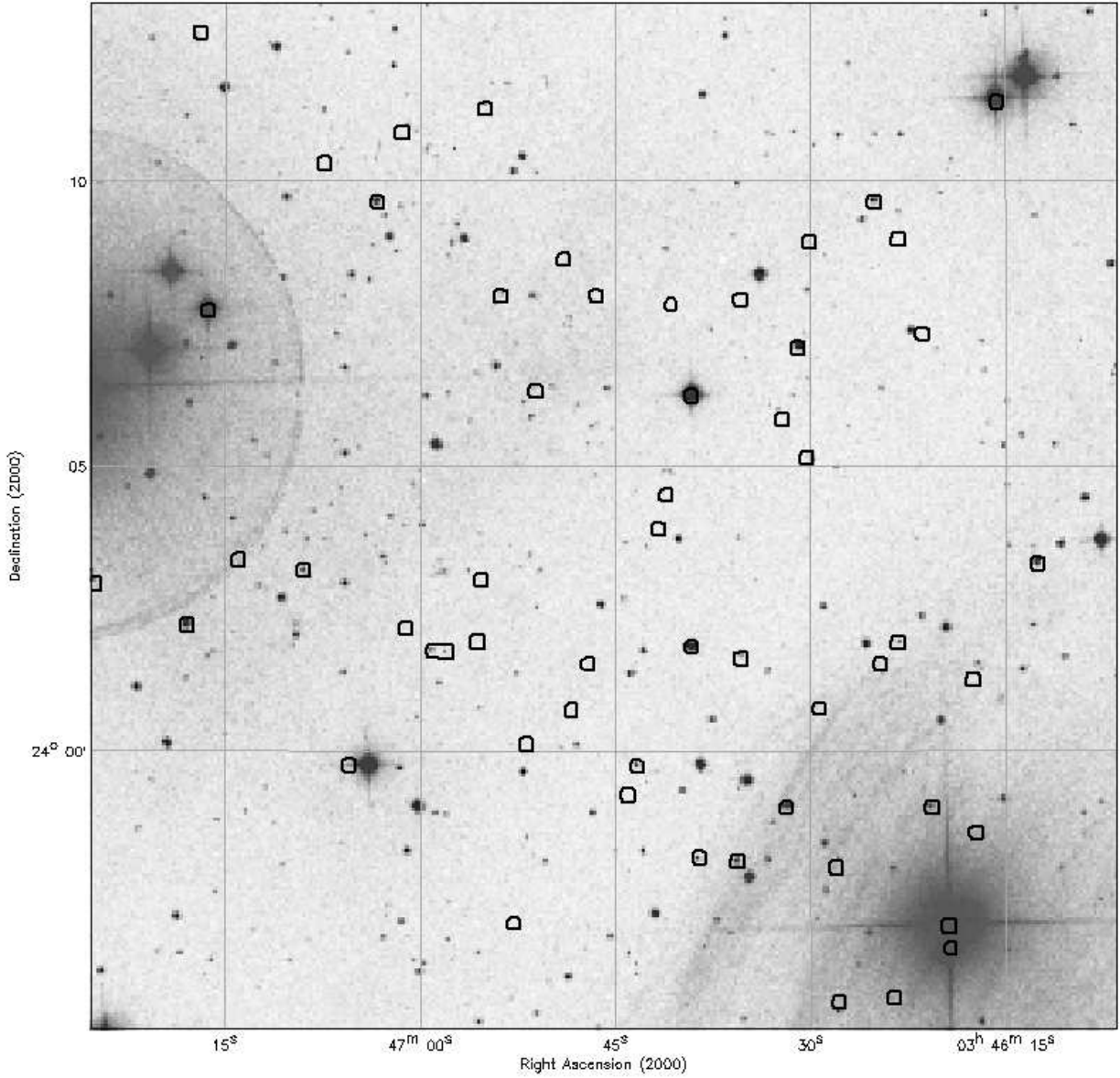,width=1.0\textwidth}
}  
\caption{ACIS-I detections of X-ray sources overplotted on an optical
image of the Pleiades field of view as seen by Chandra.  The detections
were obtained by running the ``wavdetect'' algorithm in CIAO, with the
false alarm probability set to 10$^{-6}$ per pixel.  This implies that
there would be 1 spurious source detected per 1000$\times$1000 pixel
image.  Our binned image was 512$\times$512 pixels, so we would expect
0.25 spurious sources in our field.  57 sources were using this method.}
\end{figure}

\begin{figure}
\centerline{
\psfig{figure=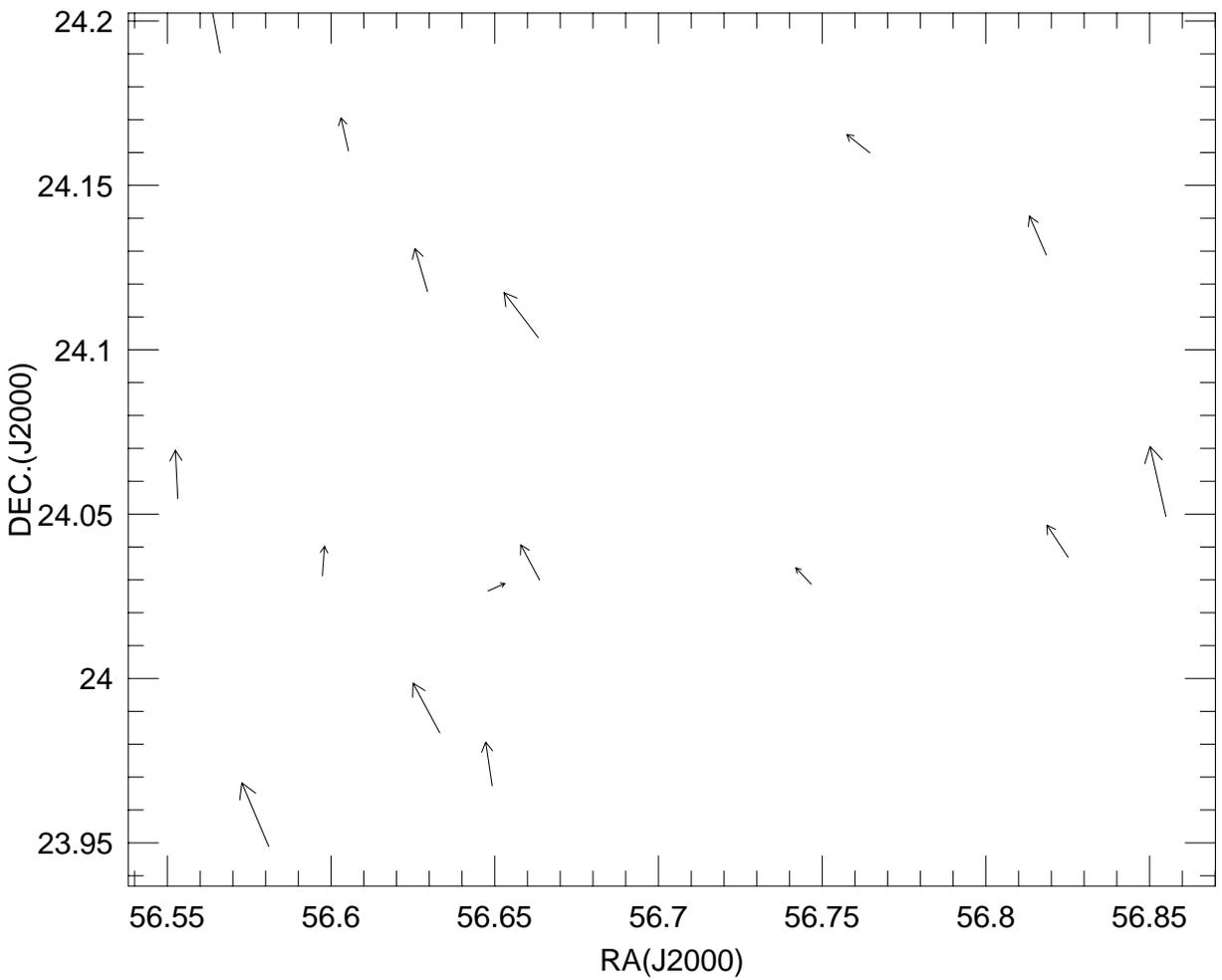,width=1.0\textwidth}
}  
\caption{Offsets between the optical counterparts and Chandra detections
before correction of the aspect solution.  The largest offset shown is
3.3$\arcsec$.}
\end{figure}

\begin{figure}
\centerline{
\psfig{figure=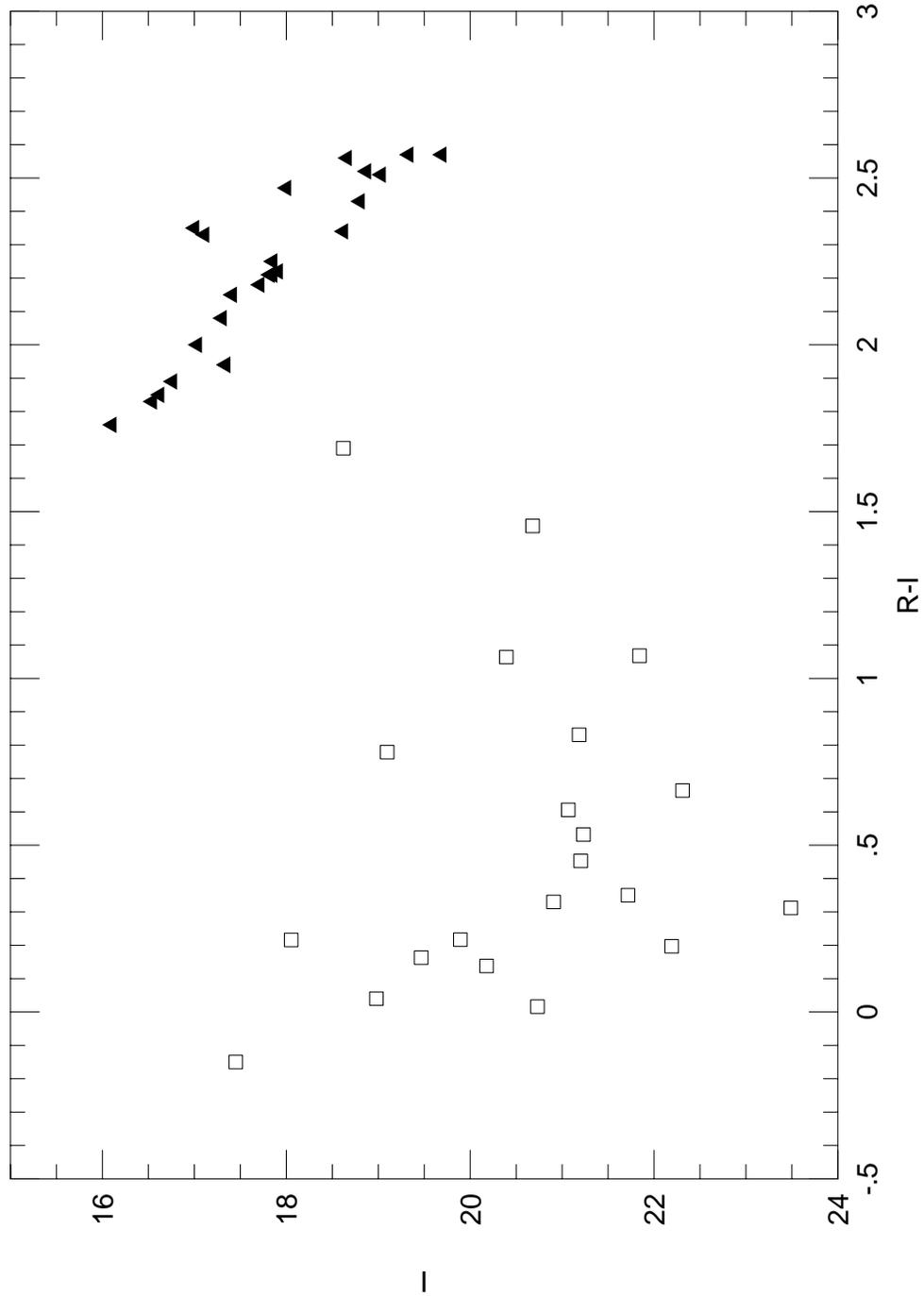,width=1.0\textwidth}
}  
\caption{I vs. R-I plot showing the Keck photometry for some of the
Chandra sources (squares).  The triangles represent data for very low
mass Pleiades members from Mart\'\i n et al. (2000).  It is seen that in
general, these Chandra detections are too blue to be low mass Pleiades
members.}
\end{figure}

\begin{figure}
\centerline{
\psfig{figure=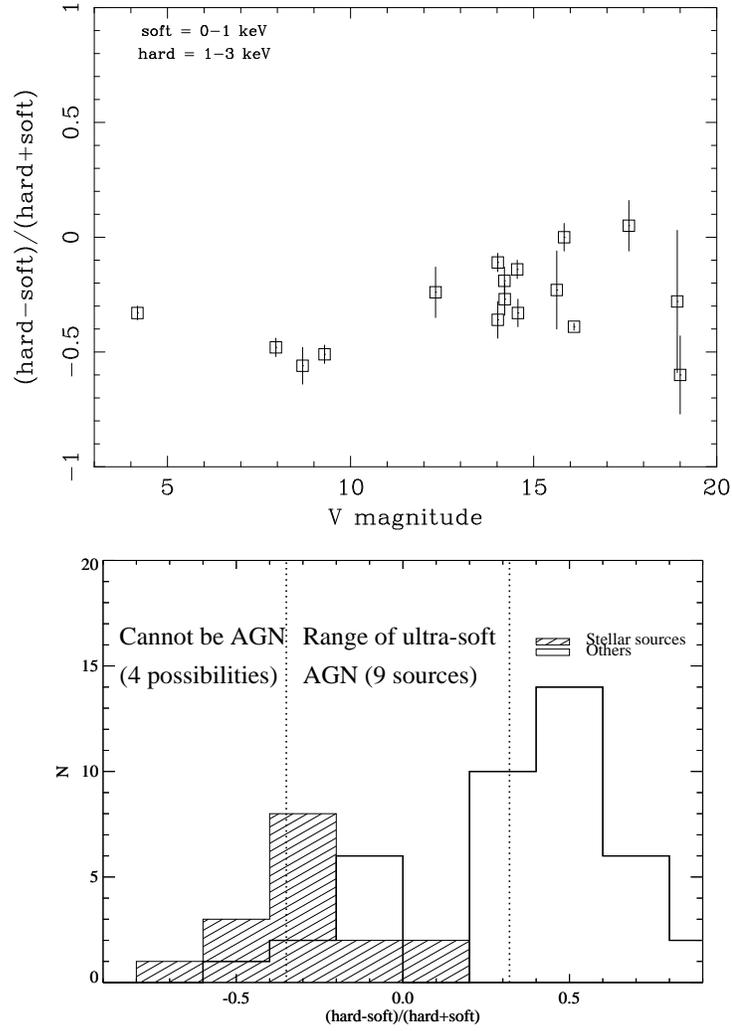,width=1.0\textwidth}
}
\caption{(a): hardness ratio vs. V magnitude for stars with known
optical counterparts.  It is seen that all of these sources are soft
although there is a tendency for bright members to be softer (b):
histogram of hardness ratios for all Chandra detections.}
\end{figure}

\begin{figure}
\centerline{
\psfig{figure=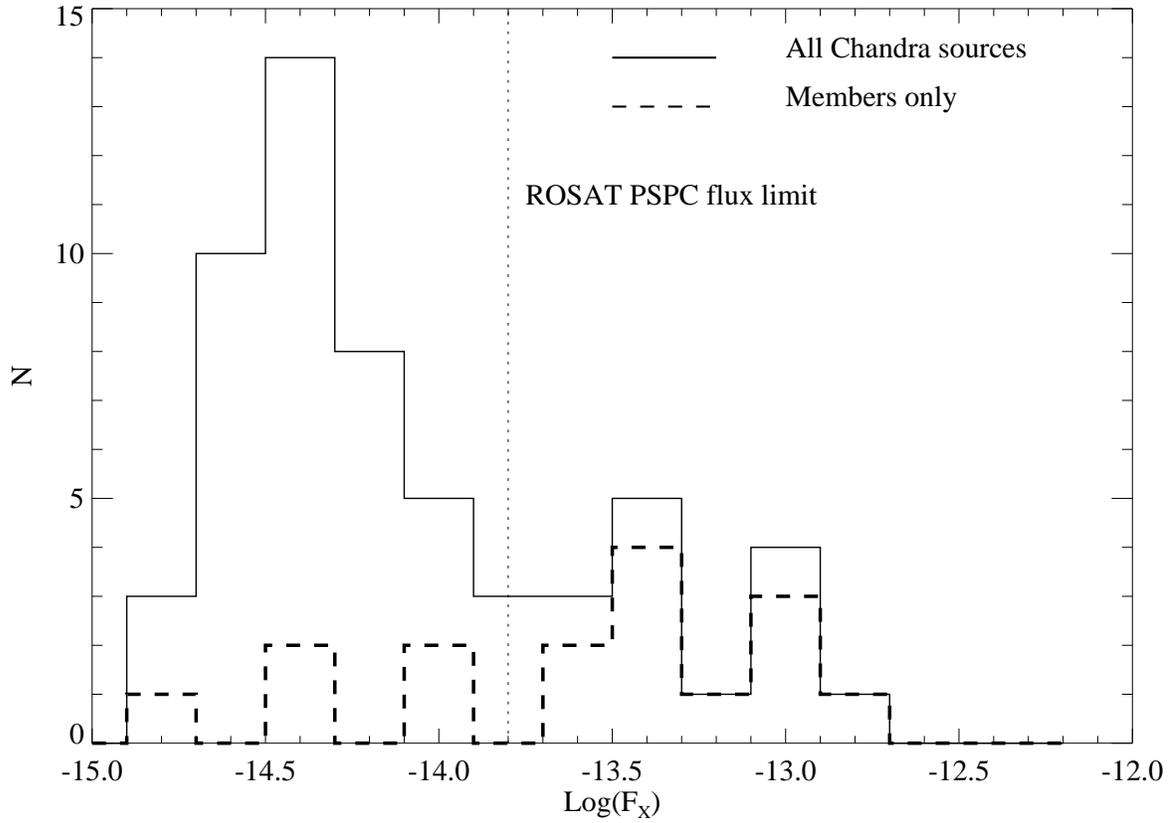,width=1.0\textwidth}
} 
\caption{Histogram of fluxes for Chandra detections in the Pleiades.
The dashed lines show the fluxes for the Pleiades members only while the
solid line shows the fluxes for all sources in the field.  Note that
most of the detections are at the fainter end of the distribution.  This
is an order of magnitude fainter than ROSAT -- the dotted vertical line
indicates the ROSAT PSPC detection limit.}
\end{figure}

\begin{figure}
\centerline{
\psfig{figure=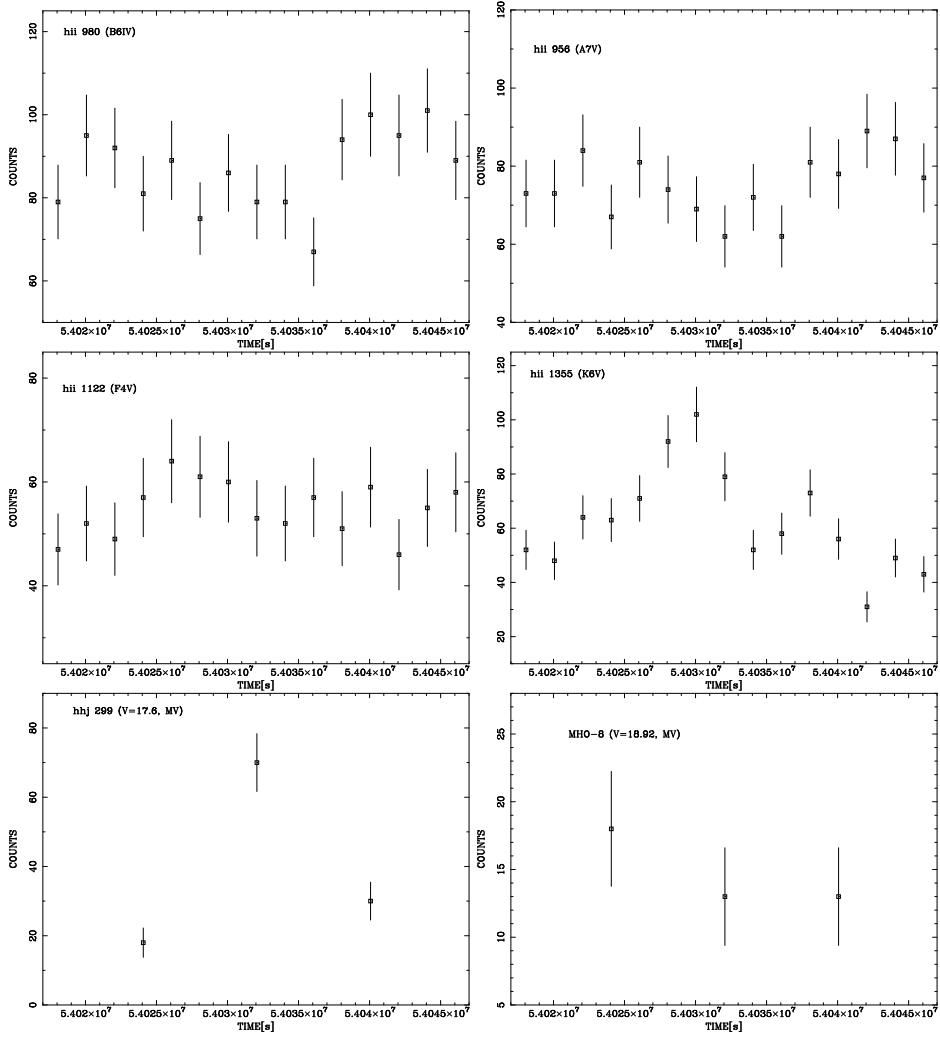,width=1.0\textwidth}
}  
\caption{Light curves for a sample of Chandra detections: hii 980
(B6IV), hii 956 (A7V), hii 1122 (F4V), hii 1355 (K6V), hhj 299 (MV),
MHO-8 (MV).}
\end{figure}

\end{document}